\documentclass[aps,prl,twocolumn,showpacs,floatfix,superscriptaddress,nofootinbib,longbibliography]{revtex4-1}
\usepackage[utf8]{inputenc}

\usepackage{amsmath}
\usepackage{amssymb}
\usepackage{inputenc}
\usepackage{amssymb}
\usepackage{graphicx}
\usepackage{epsfig}
\usepackage{bm}
\usepackage{color}
\usepackage{float}
\usepackage{dcolumn}
\usepackage{multirow} 
\usepackage{MnSymbol}
\usepackage{ulem}
\usepackage[unicode=true,
  linktocpage,
  linkbordercolor={0.5 0.5 1},
  citebordercolor={0.5 1 0.5},
  colorlinks=true,
  linkcolor=blue,
  citecolor=blue,
  urlcolor=blue]{hyperref} 
\usepackage{lipsum} 
\usepackage{placeins} 
\usepackage[dvipsnames,usenames]{xcolor}

\newcommand{\be}{\begin{equation}}
\newcommand{\ee}{\end{equation}}
\newcommand{\ba}{\begin{eqnarray}}
\newcommand{\ea}{\end{eqnarray}}

\begin{document}
\title{Spin response of neutron matter in {\it ab initio} approach}
\date{\today}
\author{J.~E.~Sobczyk}
\email{jsobczyk@uni-mainz.de}
\affiliation{Institut f\"ur Kernphysik and PRISMA$^+$ Cluster of Excellence, Johannes Gutenberg-Universit\"at, 55128
  Mainz, Germany}
\author{W.~Jiang}
\email{wjiang@uni-mainz.de}
\affiliation{Institut f\"ur Kernphysik and PRISMA$^+$ Cluster of Excellence, Johannes Gutenberg-Universit\"at, 55128
  Mainz, Germany}  

\author{A.~Roggero}
\email{a.roggerog@unitn.it}
\affiliation{Physics Department, University of Trento, Via Sommarive 14, I-38123 Trento, Italy}
\affiliation{INFN-TIFPA Trento Institute of Fundamental Physics and Applications,  Trento, Italy}
\affiliation{InQubator for Quantum Simulation (IQuS), Department of Physics, University of Washington, Seattle, WA 98195, USA}

\begin{abstract}

We propose a general method embedded in the {\it ab initio} nuclear framework to reconstruct linear response functions and calculate sum rules. Within our approach, based on the Gaussian integral transform, we consistently treat the groundstate and the excited spectrum. Crucially, the method allows for a robust uncertainty estimation of the spectral reconstruction. We showcase it for the spin response in neutron matter. Our calculations are performed using state-of-the-art many-body coupled-cluster method and Hamiltonians derived in the chiral effective field theory, emphasizing the analysis of finite-size effects. This work serves as a stepping stone towards further studies of neutrino interactions in astrophysical environments from first principles.
    
\end{abstract}

\maketitle

{\it Introduction --- }
Neutrino interactions in nuclear matter play an essential role in our understanding of dynamical processes in astrophysical environments such as neutron stars or supernova~\cite{wanajo2014production,mueller2019,Burrows_2021,Cusinato_2022,Foucart_2023,Espino2024}. Since neutrinos interact weakly, the interaction rate is directly associated with the linear response of the system. At low neutrino energies and near the saturation density where nucleons are nonrelativistic,
the response is primarily driven by the axial part of the electroweak current, which, in this regime, corresponds to spin fluctuation~\cite{Friman1979,Iwamoto1982,Raffelt1996,Reddy1999,Lykasov2008}. 

Within the domain of low momentum transfer, it is known that the response is highly sensitive to nuclear dynamics and many-body correlations in the ground state. 
Specifically, the spin response in neutron matter directly depends on the non-central terms of nuclear force, including the spin-orbit and tensor interactions. 
In the long-wavelength limit, this response arises due to the non-zero commutator between the spin operator and the nuclear Hamiltonian~\cite{Raffelt1995,Sigl:1995ac}. 

Various approaches have been proposed in the past to study the propagation of neutrinos in dense matter. Several theoretical frameworks confirm the important role played by the in-medium many-body correlations~\cite{Shen:2012sa,Lovato:2013dva,Davesne2015,Riz_2020}. For instance, the authors of~\cite{Shen:2012sa} used a Quantum Monte Carlo method~\cite{Carlson2015} and a realistic AV8 potential to reconstruct the dynamical spin response from three energy-weighted sum rules. Additionally, approximations can be made to construct the responses at low fugacity, low density, and/or high temperatures (see, e.g.,~\cite{Bedaque2018}) with designated constraints.
However, obtaining actual responses using \textit{ab initio} approaches, along with corresponding error estimation, remains an exceedingly challenging task that has yet to be achieved.

In this Letter, we present the first consistent {\it ab initio} calculation of the dynamical spin response and sum rules in pure neutron matter using Hamiltonians derived from the chiral effective field theory ($\chi$EFT). 
By doing so, we establish the foundations of a theoretical framework suitable for further studies of neutrino interactions in nuclear matter.
In the spirit of {\it ab initio}, we follow an approach which allows for a robust assessment of uncertainties coming from the many-body method and nuclear dynamics. 
We notice that significant attention has been devoted to quantify uncertainties stemming from truncations both in many-body methods and in the $\chi$EFT order by order expansion~\cite{Drischler:2020hwi, Hu:2021trw,Acharya:2022drl, Sun:2023nzj}. 
The calculation of dynamical responses, however, is a notoriously complicated problem on its own. 
At relatively large momentum transfers, the factorization schemes like short-time approximation or spectral functions offer an approximated approach to account for nuclear correlations in the prediction of the response function~\cite{Pastore2020,Andreoli2022,Rocco:2018vbf,Sobczyk:2023mey}.
For finite nuclei, the Lorentz and Laplace integral transforms were employed to access the excited spectrum through an inversion procedure~\cite{efros1994,Bacca:2014rta}. 
Although these methods have produced highly accurate results in the quasi-elastic peak regime~\cite{Lovato:2016gkq,Sobczyk:2021dwm,Sobczyk:2023sxh}, they rely on the smoothing assumption which introduces uncontrolled errors. 
Here, we present a versatile method based on the Gaussian integral transform (GIT) which circumvents the inversion procedure allowing for the rigorous uncertainty estimation~\cite{Roggero:2020qoz,Sobczyk:2021ejs}. 
It opens the door for further studies in which all the relevant sources of uncertainty in a many-body calculation can be systematically included.

{\it Methods --- }
We propose an {\it ab initio} framework to reconstruct a general observable which is expressed in terms of an integral with a linear response function $S(\omega)$
\begin{equation}
    Q(S,f) = \int d\omega S(\omega) f(\omega)\, .
    \label{eq:observable}
\end{equation}
In this work we will consider $f(\omega)$ to be the energy-weighted sum rules, as well as the window function. We note, however, that it can be any bound function.
The linear response of an infinite system excited with an operator $\theta$,
\begin{equation}
    S(\omega) = \int d\omega |\langle f |\theta |0\rangle|^2\delta(E_0+\omega-E_f),
    \label{eq:response}
\end{equation}
will be reconstructed in terms of the GIT.
In our approach we express a general integral transform of the response function on a basis of orthogonal polynomials $T_k$, leading to
\begin{equation}
    R(\omega) = \int d\omega' K_\lambda(\omega,\omega') S(\omega') = \sum_k^\infty c_k(\omega,\lambda) m_k
\end{equation}
with an integral kernel $K_\lambda$ taken to be the Gaussian with variance $\lambda$ and with the moments of expansion $m_k = \int d\omega S(\omega) T_k(\omega)$. The width of the integral kernel controls the energy resolution.
The observable of interest in Eq.~\eqref{eq:observable} can be approximated using the GIT
\begin{align}
    Q(R,f) =& \int d\omega R(\omega) f(\omega) = \\ \nonumber
    &\int d\omega \left(\int d\omega' K_\lambda(\omega,\omega') f(\omega') \right) S(\omega),
\end{align}
where the error of reconstruction $| Q(R,f) -  Q(S,f)|$ can be rigorously estimated directly from the form of the kernel and number of moments $m_k$ used to expand it. 
In particular, the response function $S(\omega)$ can be reconstructed in terms of a histogram, estimating the errors of spectral reconstruction, following Ref.~\cite{Sobczyk:2021ejs}, avoiding this way an uncontrollable inversion procedure.
For these applications, the Gaussian kernel is much better suited than the Lorentzian, leading to faster convergence and much stricter error bounds~\cite{Sobczyk:2021ejs}. The choice of orthogonal polynomials employed in the expansion depends on the approach used to estimate moments, for simulations on quantum computers both Fourier~\cite{Roggero2019,Somma_2019,Hartse2023,Kiss:2024foh} and Chebyshev~\cite{Roggero:2020qoz,Roggero2020,Summer2024} moments could be used, for classical many-body methods the latter are instead preferred.

We apply the coupled-cluster (CC) theory for nuclear matter, as introduced in Ref.~\cite{Hagen:2013yba}, to perform our calculations. It has been successfully applied in previous studies to learn the properties of both nuclear matter and pure neutron matter ~\cite{Ekstrom:2017koy,Jiang:2020the,Gu2023}. The approach is based on the similarity transformed Hamiltonian $ \overline{H} = e^{-T}H e^{T}$ where $T$ is the cluster operator that induces $n$ particle $n$ hole correlations. Since $\overline{H}$ is non-Hermitian, the left and right ground states are parameterized differently: 
\begin{equation}
\langle \widetilde \Psi_0| = \langle 0| (1+\Lambda)e^{-T},~~|\Psi_0\rangle=e^T |0\rangle
\end{equation}
with $\Lambda$ as de-excitation operator and $|0\rangle$ the closed-shell Fermi vacuum. In this paper, all CC results are obtained within the coupled-cluster doubles (CCD) approximation, i.e., 
$T =  \frac{1}{4}\sum_{ijab} t^{ab}_{ij} a_a^\dagger a_b^\dagger a_i a_j $ and $\Lambda =  \frac{1}{4}\sum_{ijab} \lambda^{ij}_{ab}  a_i^\dagger a_j^\dagger a_a a_b$. Here, the amplitudes $t(\lambda)$ are obtained by solving a set of coupled nonlinear equations. The many-body system is solved on a cubic lattice in momentum space using twisted angle boundary conditions (TABC)~\cite{Hagen:2013yba}. The model space has $(2n_{\rm max} + 1)^3$ momentum points. We employ $n_{\rm max}=3$ for which our results are well converged. It is important to note that for the pure neutron matter system under investigation, CCD serves as a good approximation, yielding a small truncation error in CC expansion. Specifically, the error is $\sim 5\%$ of the correlation energy, translating to around 0.04 MeV for the ground-state energy at saturation density~\cite{Hu:2021trw}. 

To study the spin response in nuclear matter, we needed to substantially extend the existing implementation of the CC method by introducing the similarity-transformed excitation operators $\overline\theta \equiv e^{-T}\theta e^{T}$ and the equation-of-motion (EOM) technique~\cite{Stanton:1993vcu}.
The principle idea of EOM is that a target state $|\Phi\rangle$, such as $|f\rangle$ in Eq.~\eqref{eq:response}, can be generated from the initial state $|0\rangle$ by $|\Phi\rangle= \mathcal{R}|\Psi_0 \rangle$, where $\mathcal{R} =  r_0+ \frac{1}{4}\sum_{ijab} r^{ab}_{ij} a_a^\dagger a_b^\dagger a_i a_j$ including all possible excitations in present CC truncation. Since the CC theory is non-Hermitian, the left target states have to be determined separately following the ansatz $\langle  \widetilde \Phi| = \langle \widetilde \Psi_0| \mathcal{L}$ with $\mathcal{L}=l_0+ \frac{1}{4}\sum_{ijab} l^{ij}_{ab} a_i^\dagger a_j^\dagger a_a a_b$.

With these ingredients, we are able to employ the EOM to retrieve the Chebyshev moments of the GIT expansion, $m_k$, following recursive relations for the Chebyshev polynomials. Utilizing the solved $T$ and $\Lambda$ of the ground state, the starting pivots can be written as
\begin{equation}
\langle \widetilde \Phi_0| = \langle 0| (1+\Lambda) \overline \theta^\dagger\, , \ \ \ \ \ |\Phi_0\rangle = \overline \theta |0\rangle.
\end{equation}
Then we iteratively calculate states $\Phi_k$ and corresponding $m_k$,
\begin{align}
     \langle \widetilde \Phi_k| &= \langle \widetilde \Phi_{k-1}| \overline{H}_{\rm norm}\, ,  &|\Phi_k\rangle &= \overline{H}_{\rm norm} |\Phi_{k-1}\rangle \nonumber \\ 
    m_0 &= \langle \widetilde \Phi_0|\Phi_0\rangle \, ,  &m_1 &=  \langle \widetilde \Phi_0|\Phi_1\rangle \nonumber \\ 
    m_{k+1} &= 2 \langle \widetilde \Phi_0|\Phi_{k+1}\rangle - m_{k-1}& \equiv 2& \langle \widetilde \Phi_{k+1}|\Phi_{0}\rangle - m_{k-1}\,.
\end{align}
This expansion requires the Hamiltonian spectrum to be contained in the range $(-1,1)$ in which the Chebyshev polynamials are defined. In order to do this, we introduce normalization factors which depend on the maximum and minimum excitation energy to properly rescale the spectrum, $a = (E_{\rm max}-E_{\rm min})/2$, $b = (E_{\rm max}+E_{\rm min})/2$ and define a normalized Hamiltonian $\overline{H}_{\rm norm} \equiv (\overline{H}-b)/a$. They depend on the model space in which we perform calculations and the Hamiltonian itself, and are typically of the order of a few-hundreds MeV.

We point out that the recursive method of obtaining moments at each step requires only two states, $\Phi_0$ and $\Phi_k$. It is numerically stable, and does not involve additional orthogonalization procedure, contrary to the established Lanczos algorithm. While the Lanczos algorithm is an excellent approach to get the approximated extreme eigenvalues with a limited number of iterations, the polynomial expansion provides a way to reconstruct the whole spectrum using an integral transform of a given resolution $\lambda$.
For finite nuclei, the LIT is usually expressed as a continuous fraction using Lanczos coefficients~\cite{Bacca:2014rta}.
Since our approach can be applied to any integral transform kernel, we benchmarked our results for the Lorentz kernel of width $\lambda=5$ MeV, using either Chebyshev expansion or Lanczos tri-diagonalization procedure which led to the same results.

{\it Spin response and sum rules ---}
The spin response $S_\sigma(\omega,q)$ is given by Eq.~\eqref{eq:response} with the following choice for the excitation operator
\begin{equation}
    \theta = V^{-1}\sum_{i=1}^N {\bm \sigma}_i \exp (i{\bf q}\cdot {\bf r}_i)
\end{equation}
and we introduce a global factor $4/3\rho$ with $\rho=N/V$ being the neutron number density, to account for the canonical normalization $S_\sigma(q\to \infty) = 1$.
Currently we investigate the response in the long-wavelength limit $q\to 0$, however, the approach we present here can be directly extended to the calculation of the response for finite values of the momentum transfer.
In the first step, we calculate the total strength $Q^0_\sigma$ and the energy-weighted sum rule $Q^1_\sigma$ where the general energy-weighted sum rules are defined as
\begin{equation}
    Q^n_\sigma = \int d\omega S_\sigma(\omega) \omega^n\, .
\end{equation}
We note that $Q^0_\sigma$ can be calculated in three ways {\it (i)} as an expectation value of the two-body operator $\sum_{i,j}\bm{\sigma}_i \bm{\sigma}_j$, {\it (ii)} as the first moment of Chebyshev expansion, and {\it (iii)} as an integral of the discretized response function. Similarly for the $Q^1_\sigma$ sum rule, it can be calculated {\it (i)} from the first two moments of the Chebyshev expansion or {\it (ii)} as an expectation value of the tensor part of the Hamiltonian, following the derivation of Ref.~\cite{Sigl:1995ac}.  We performed distinct calculations which led to the same numerical results for both sum rules within the CCD framework.

To assess the sensitivity of the results to the details of the many-body expansion, we performed a comparison between two distinct methods, using a simple LO chiral Hamiltonian from Ref.~\cite{Ekstrom:2017koy} and we benchmarked the results obtained with the CCD method against the configuration-interaction Monte Carlo (CIMC)~\cite{Roggero2014,Arthuis2023}.
The calculations of two different many-body methods are carried out in the same basis setup and an equal number of particles in the system. For binding energies, the deviations between the two methods are within $0.5\%$. For sum rules and average excitation energy $\omega_1$$=Q_\sigma^1/Q_\sigma^0$, the differences are $\sim 20\%$ considering the different many-body correlations adopted by the two methods.

In Table~\ref{table:sum_rules} we collect numerical values for the sum rules at three densities, employing the chiral Hamiltonian NNLO$\Delta_{\rm GO}(394)$ and the model space of $n_{max}=3$ for $N=66$, 114 particles. We assess the uncertainties coming from the finite size effects by performing calculations with twisted angle boundary conditions (TABC) for three values of angle taken in the Gaussian nodes, following the strategy in Ref.~\cite{Hagen:2013yba}. This amounts to 27 TABC setups. In Table~\ref{table:sum_rules} we report the mean value and $68\%$ uncertainty. Results for $N=66$ and $N=114$ stay in agreement, the latter being systematically few percent lower. The uncertainty of $N=114$ is smaller and exhibits a mild $\rho$ dependence. This can be expected, since the quantized nucleon momenta depend $\sim(\rho/N)^{1/3}$.
We note that our results substantially differ from the values reported in Ref.~\cite{Shen:2012sa}. The total strength is around five times lower. Moreover, the average excitation energy $\omega_1$ is considerably higher which suggests that the strength of the response will peak at higher energies.
We prescribe these differences mainly to the nuclear Hamiltonian, and to a smaller extent to the many-body wavefunction. Note that the delta-full nuclear interaction used in this paper is notably soft, ensuring rapid convergence as the model space increases. Furthermore, the explicit inclusion of delta isobar, i.e. excited states
of the nucleon that reflects its finite size, should in principle increase the EFT breakdown scale and provide an improved description of nuclear matter at higher density~\cite{Ekstrom:2017koy,Jiang:2020the}.

\begin{table}[t]
\renewcommand{\arraystretch}{1.5} 
	\centering
	\begin{tabular}{c c c c c} 
		\hline
$\rho$ [fm$^{-3}$] &  N &  $Q^0_\sigma$ & $Q^1_\sigma$ [MeV] & $\omega_1$ [MeV] \\ [1.5ex] 
\hline
0.08 & 66 & $0.0380 ^{+0.0015}_{-0.0015}$ & $3.344 ^{+0.111}_{-0.090}$ &$87.99^{+0.96}_{-0.75}$ \\[0.5ex] 
 & 114 & $0.0359 ^{+0.0005}_{-0.0010}$ & $2.954 ^{+0.059}_{-0.059}$ & $82.26^{+1.43}_{-0.89}$ \\[0.5ex] 
\hline
0.12 & 66 & $0.0371 ^{+0.0018}_{-0.0032}$  & $3.636 ^{+0.154}_{-0.222}$  &$98.13^{+2.30}_{-1.57}$  \\[0.5ex] 
 & 114 &  $0.0357 ^{+0.0006}_{-0.0016}$ & $3.376 ^{+0.060}_{-0.072}$ &$94.79^{+2.10}_{-0.24}$ \\[0.5ex] 
\hline
0.16 & 66 & $0.0309 ^{+0.0014}_{-0.0045}$  & $3.200 ^{+0.134}_{-0.346}$ &$104.13^{+4.20}_{-2.48}$\\[0.5ex] 
 & 114 & $0.0298 ^{+0.0008}_{-0.0019}$ & $3.011 ^{+0.065}_{-0.103}$ & $101.37^{+2.76}_{-0.90}$\\ [0.5ex] 
\hline
	\end{tabular}
	\caption{
 $Q^0_\sigma$, $Q^1_\sigma$ sum rules and $\omega_1=Q^1_\sigma/Q^0_\sigma$ calculated for $\rho=0.08,\ 0.12, \ 0.16$ fm$^{-3}$ with $N=66,\, 114$ neutrons, using NNLO$\Delta_{\rm GO}(394)$ Hamiltonian. }
	\label{table:sum_rules}
\end{table}

\begin{figure}[b]
    \includegraphics[width=0.45\textwidth]{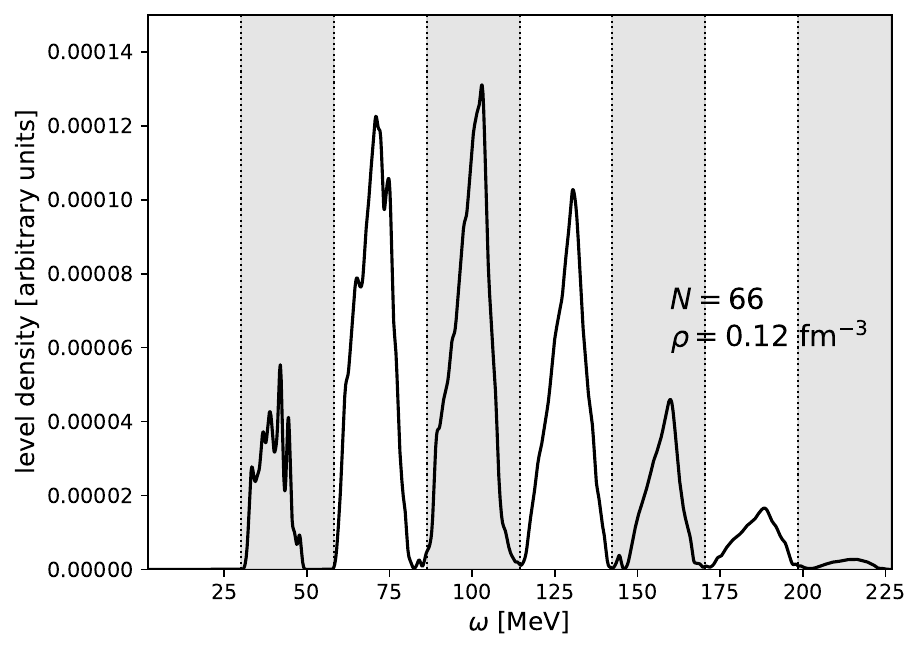}
  \caption{GIT of the level density for $\rho=0.12$ fm$^{-3}$ and $N=66$ particles. Vertical lines mark the binning driven by $E_{gap}$.}
  \label{fig:level_density}
\end{figure}

Within our approach we have access to the excited spectrum of the system which can be reconstructed in terms of histograms, using $f(\omega)$ from Eq.~\eqref{eq:observable} as the window function
\begin{equation}
\label{eq:window}
f(\omega; \eta, \Delta) = \bigg\{\begin{matrix}
0 & |\eta-\omega|>\Delta\\
1 & \text{otherwise}
\end{matrix}\;
\end{equation}
where the half-width of each bin, $\Delta$, is a free parameter which is a priori unknown and $\eta$ is the center of the bin.
In the thermodynamical limit, the continuous excited spectrum could be reconstructed to any high precision, choosing $\Delta\ll 1$. However, in the discretized space this is not the case. To inspect the details of the discretized continuous spectrum, we first investigated the density of states, randomizing the excitation operator $\theta$ from Eq.~\eqref{eq:response}. We perform the GIT expansion into Chebyshev polynomials using a high resolution kernel (i.e. setting the Gaussian width $\lambda$ to a small value).
As an example, we show in Fig.~\ref{fig:level_density} the spectrum for the case of $N=66$ particles at $\rho=0.12$ fm$^{-3}$, using GIT with $\lambda=0.5$ MeV kernel and 5000 Chebyshev moments.
The excitation spectrum has a very distinct irregular shape with the first excited state lying $\sim 30$ MeV above the groundstate. This position can be estimated following an intuitive argument. The groundstate energy of the system is driven by the mean-field value, $i.e.$ all the particles occupy the lowest orbitals. The minimal energy required to excite the system, involves exciting two particles lying just below the Fermi level to the lowest possible state above (to conserve the total momentum). We calculated the position of the first excited state using the EOM for  $N=14$, 38, 54, 66, 114 particles. These numerical results agree within $5-30\%$ with the kinetic energy difference between the last occupied shell and the first unoccupied shell which scales as $\sim (N/\rho)^{-2/3}$. This allows us to understand the sensitivity of our method to the lower part of the spectrum which is limited due to discretization. 
Moreover, the pattern of the excited spectrum can be understood in terms of the employed underlying shell structure. The kinetic energy is quantized, and the energy required to excite two particles from a lower shell to a higher one is given by $E_n = 2n E_{gap}$ with $E_{gap} = \hbar^2/2m (2\pi/L)^2$ where $L=(N/\rho)^{1/3}$,  $n=1,2,...$. 
The excited states are clustered in between these ``shell closures'' leading to a high density in these regions.
The values of energy $E_{gap}$ agrees with the numerical values we found with the GIT. 
Following these observations, we set the binning to reconstruct the discretized spectrum in terms of a histogram. At the density $\rho=0.12$ fm$^{-3}$ the binning corresponds to $2\Delta=75$, 40, 28, 20 MeV for $N=14$, 38, 66 and 114 particles respectively. 
\begin{figure}[hbt]
    \includegraphics[width=0.45\textwidth]{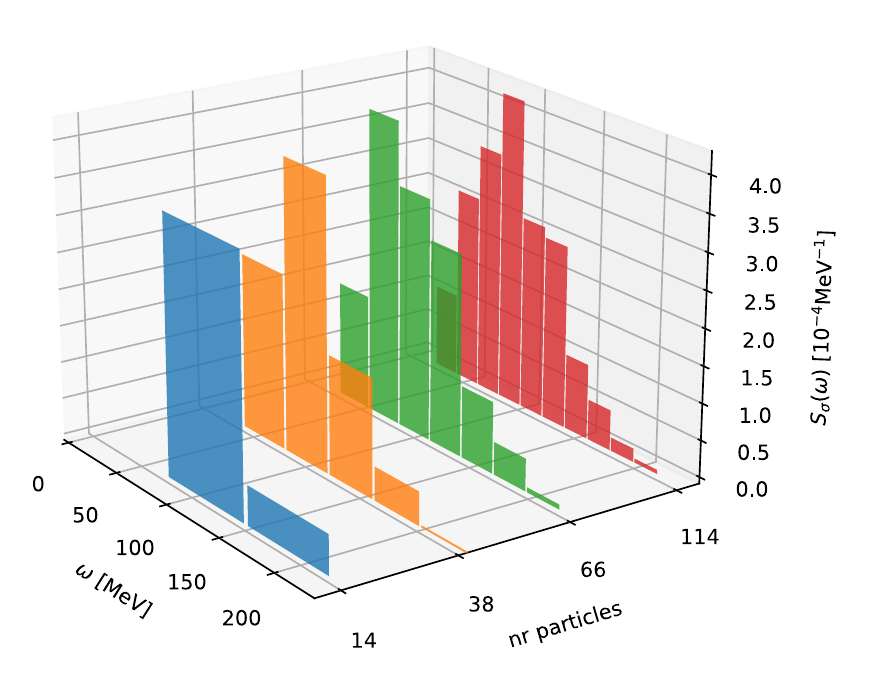}
  \caption{Spin response at $\rho=0.12$ fm$^{-3}$ for $N=14,38,66,114$ neutrons reconstructed in terms of histograms.}
  \label{fig:response_np}
\end{figure}

The analysis performed for the level density is general and the same binning can be directly applied to the spectral reconstruction of the spin response. In Fig.~\ref{fig:response_np} we show the results for an increasing number of particles used in the simulation. For the histogram reconstruction we performed calculations using TABC and took the mean value, similarly to the calculations of the sum rules presented earlier. As we pointed out and showed in Table~\ref{table:sum_rules}, the sum rules $Q^0_\sigma$ and $Q^1_\sigma$ do not vary much with $N$, i.e. the position of the peak $\omega_1\approx Q^1_\sigma/Q^0_\sigma$ is similar within all the calculations. However, the details of the discretized excited spectrum and sensitivity to its low energy part depend strongly on the number of particles.

\begin{figure*}[hbt]
    \includegraphics[width=0.328\textwidth]{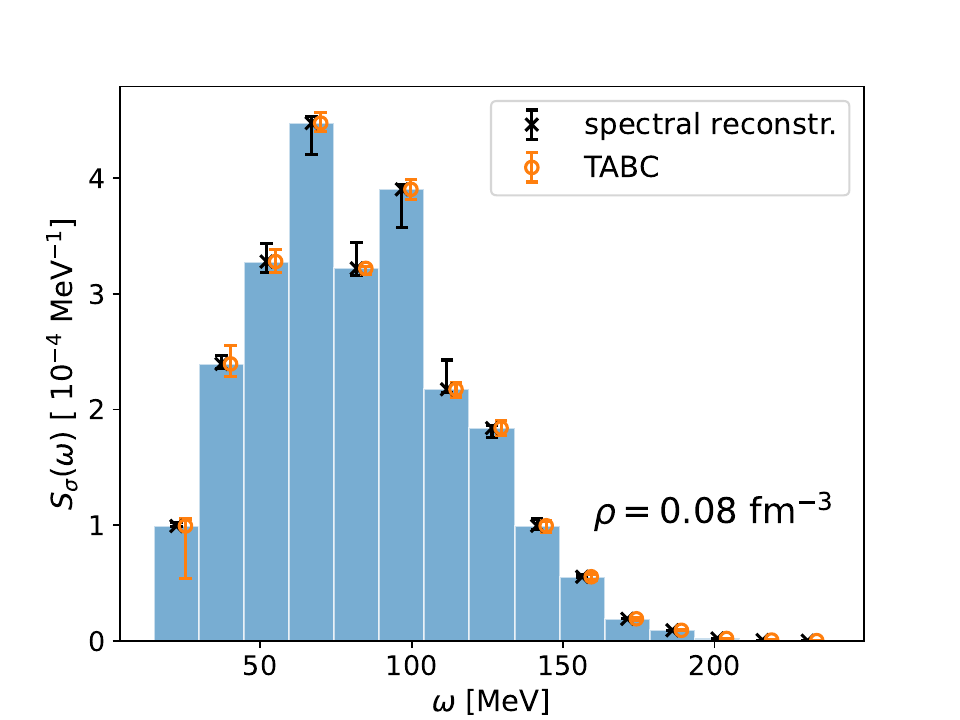}
    \includegraphics[width=0.328\textwidth]{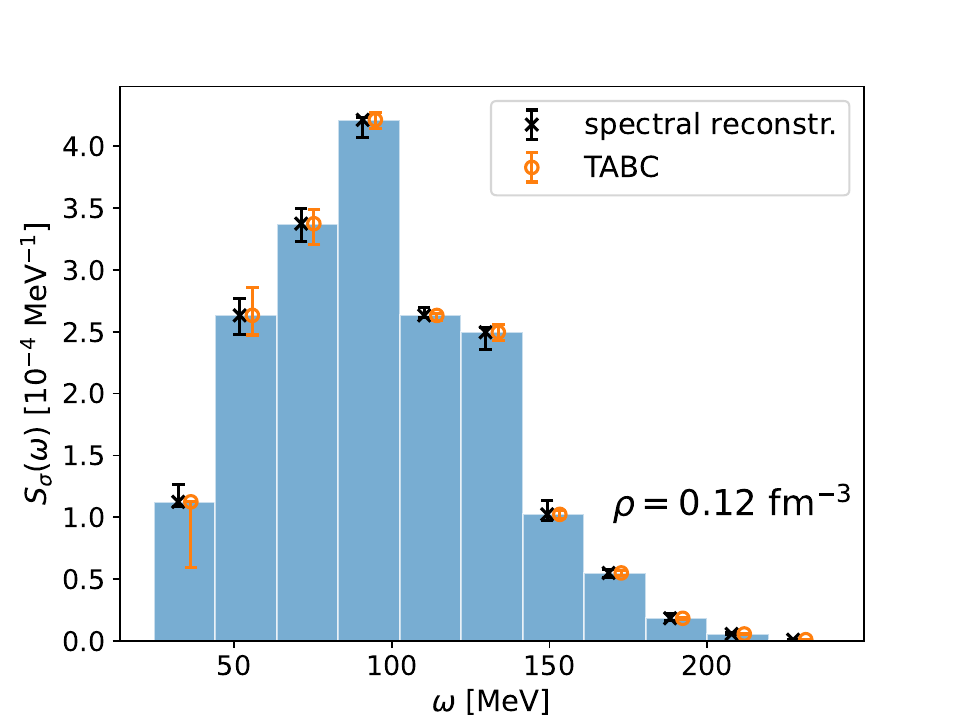}
    \includegraphics[width=0.328\textwidth]{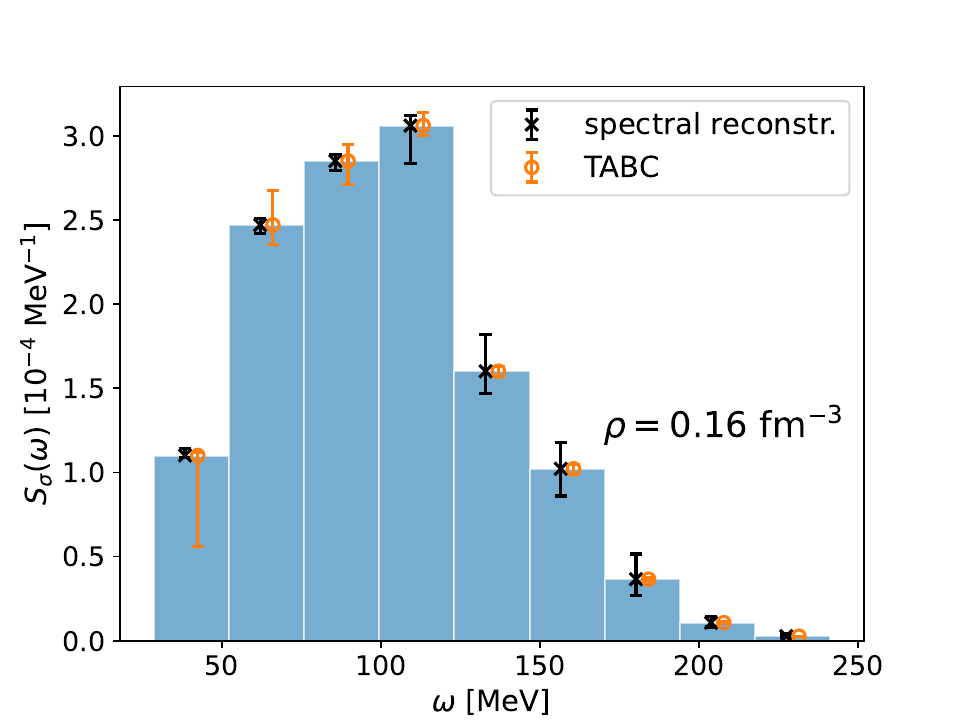}
    
  \caption{Spin response at densities $\rho=0.08,\ 0.12,\ 0.16$ fm$^{-3}$ for $N=114$ particles reconstructed in terms of histograms.}
  \label{fig:response_114}
\end{figure*}

Simulations using 114 particles allow us to reconstruct the response in bins of 15 MeV for $\rho=0.08$ fm$^{-3}$, up to 24 MeV for $\rho=0.16$ fm$^{-3}$. In Fig.~\ref{fig:response_114} we show these results, together with the error estimation coming from two sources related to the fact that our calculations are performed in a discretized space.
The first source of error is related to our procedure of constructing a histogram from a discretized spectrum. We follow here closely the error estimation for GIT which we derived in Ref.~\cite{Sobczyk:2021ejs} and explained in more detail in the supplemental material. It depends on the type of the kernel, its width $\lambda$, the number of Chebyshev moments, and the size of the bins. 
This error could be further reduced by applying a higher resolution kernel which translates into the need of calculating more Chebyshev moments, $i.e.$ higher computational cost. 
Secondly, we estimate the uncertainty coming from the final size effects, using TABC for three twist angles. Similarly to the sum rule estimation, we report the mean value and standard deviation. As can be observed in Fig.~\ref{fig:response_114}, the error is dominating at the lower part of the spectrum. 

{\it Summary and outlook --- }
We have conducted the first consistent calculation of the spin response in pure neutron matter employing the nuclear {\it ab initio} approach. Our framework aims to consistently reconstruct the excitation spectrum, with particular attention to error estimation, an aspect that has not been fully explored in the past.
In this first calculation, we did not attempt to estimate the systematic uncertainty coming from the choice of the nuclear interaction. We expect this to be the dominant residual source of systematic error in our present calculations and are planning to provide a direct estimation in follow-up work. We note that besides performing simulations with different Hamiltonians, the calculation strategy laid out in the present work will remain unchanged.
We believe that the present framework holds significant potential for advancing studies of neutrino interactions in astrophysical environments and finding possible applications in other complex many-body systems.

\vspace{0.3cm}
\begin{acknowledgements}
We would like to acknowledge useful discussions with J. Carlson and S. Gandolfi about the results of this work. 
The present research is supported by the European Union’s Horizon 2020 research and innovation programme under the Marie Skłodowska-Curie grant agreement No.~101026014 (J.~E.~S.), by the Deutsche Forschungsgemeinschaft (DFG) through the Cluster of Excellence ``Precision Physics, Fundamental Interactions, and Structure of Matter" (PRISMA$^+$ EXC 2118/1) funded by the DFG within the German Excellence Strategy (Project ID 39083149) (J.~E.~S.). Computer time was provided by the supercomputer MogonII at Johannes Gutenberg-Universit\"{a}t Mainz. A.R. acknowledges support from ICSC – Centro Nazionale di Ricerca in HPC, Big Data and Quantum Computing, funded by the European Union under NextGenerationEU. Views and opinions expressed are however those of the author(s) only and do not necessarily reflect those of the European Union or The European Research Executive Agency. Neither the European Union nor the granting authority can be held responsible for them.
\end{acknowledgements}

\bibliography{biblio}

\begin{widetext}
\section{Supplemental material}
\subsection{Error bounds on histogram reconstruction}
When reconstructing a dynamical response $S(\omega)$ in terms of a histogram, we define a single bin of half-width $\Delta$ centered at $\eta$  by the window function $f(\omega,\eta;\Delta)$ (see Eq.~\eqref{eq:window} in the main text). Let us denote the total strength $Q^0 = \int d\omega\, S(\omega)$.
We are interested in approximating 
\begin{equation}
    h(\eta;\Delta) = \int_{E_{\rm min
}}^{E_{\rm max}} d\omega S(\omega)f(\omega,\eta;\Delta) = \int_{\eta-\Delta}^{\eta+\Delta}d\omega S(\omega)\;
    \label{eq:hist_bin}
\end{equation}
using an integral transform
\begin{equation}
\label{eq:hist_trans}
\widetilde{h}(\eta;\Delta) =\int_{E_{\rm min
}}^{E_{\rm max}} d\omega  \int_{\eta-\Delta}^{\eta+\Delta} d\nu K(\nu, \omega) S(\omega)\;,
\end{equation}
where $K$ is a general kernel which we define to be $\Sigma$--\textit{accurate} with $\Lambda$--\textit{resolution} if
\begin{equation}
\inf_{\omega_0\in (E_{\rm min}, E_{\rm max})}\int_{\omega_0-\Lambda}^{\omega_0+\Lambda} d\nu K(\nu, \omega_0) \geq 1-\Sigma\;.
\label{eq:kernel_definition}
\end{equation}
In this work we take its specific realization as a Gaussian with variance $\lambda$ which yields 
\begin{equation}
    \Sigma \leq \exp\big( -\frac{\Lambda^2}{2\lambda^2}  \big)\,.
    \label{eq:sigmaG}
\end{equation}
The integral transform from Eq.~\eqref{eq:hist_trans} can be approximately calculated by expanding the kernel on a basis of orthogonal polynomials and retrieving the moments of expansion from the many-body method.

In Ref.~\cite{Sobczyk:2021ejs} we derived the analytical expressions to bound the reconstruction error. All the details are presented in Appendices B, C of the latter work. The uncertainty is comprised of two sources. First of all, we use a $\Sigma$--\textit{accurate} kernel with $\Lambda$--\textit{resolution} which allows us to bound the strength in each bin as (see App.~C of~\cite{Sobczyk:2021ejs})
\begin{equation}
\widetilde{h}^\Lambda(\eta;\Delta-\Lambda) - Q^0 \Sigma \leq h(\eta,\Delta) \\
 \leq \widetilde{h}^\Lambda(\eta;\Delta+\Lambda) + Q^0 \Sigma\;.
\label{eq:histogram_bound}
\end{equation}
Secondly, the integral transform of Eq.~\eqref{eq:hist_trans} is approximated by a finite number of moments $M$. This truncation can be included to the error bound leading to the final result
\begin{equation}
\widetilde{h}^\Lambda_M(\eta;\Delta-\Lambda) - Q^0 \Sigma-2 Q^0 \beta_M\left(\Delta-\Lambda\right) \leq h(\eta,\Delta)  \leq \widetilde{h}^\Lambda_M(\eta;\Delta+\Lambda) + Q^0 \Sigma+2 Q^0 \beta_M\left(\Delta+\Lambda\right)\;,
\label{eq:histogram}
\end{equation}
where $\beta_M$ depends on the kernel properties and variance $\lambda$. For the Gaussian it can be bound by an analytical expression reported in Eq.~(B10) of Ref.~\cite{Sobczyk:2021ejs}.

We use Eq.~\eqref{eq:histogram} to put bounds on the reconstruction of the spin response at three values for the number density, $\rho = 0.08$, $0.12$, $0.16$ fm$^{-3}$. Since we expand the GIT into Chebyshev polynomials, we need to rescale the spectrum to the interval $[-1,1]$ in which they are defined. In this work, the employed Hamiltonian and model space lead to $E_{\rm min} = 0$, $E_{\rm max} = 1200$ MeV, yielding the rescaling coefficients $a=b=600$ MeV. We employ the Gaussian of $\lambda=0.5$ MeV to realize the general kernel $K$ with $\Lambda = 2.5$ MeV. This choice leads to a sufficient suppression of $\Sigma$, given in Eq.~\eqref{eq:sigmaG}, so that it becomes a negligible contribution in the uncertainty budget of Eq.~\eqref{eq:histogram}. Moreover, we have to estimate the truncation error, driven by $\beta_M$. Analytical bounds can be found in App. B of Ref.~\cite{Sobczyk:2021ejs}. However, we note that these bounds are very conservative. Here, we follow a different strategy and we estimate this error numerically. We checked that with $M=5000$ moments the results are well converged, so to the numerical precision $\widetilde{h}^\Lambda(\eta;\Delta-\Lambda) =\widetilde{h}^\Lambda_M(\eta;\Delta-\Lambda)$. This leads to negligible
$2\beta_M\left(\Delta\pm\Lambda\right)$ and the error bound becomes
\begin{equation}
\widetilde{h}^\Lambda_M(\eta;\Delta-\Lambda) \leq h(\eta,\Delta) \leq \widetilde{h}^\Lambda_M(\eta;\Delta+\Lambda)\;.
\label{eq:histogram2}
\end{equation}
For the case of $N=114$ particles we reconstruct the response in bins of $2 \Delta = 15,\ 20,\ 24$ MeV (corresponding to $\rho=0.08,\ 0.12,\ 0.16\ {\rm fm}^{-3}$), keeping the same GIT resolution. We calculate $\widetilde{h}^\Lambda_M$ exactly using coefficient of Chebyshev expansion reported in Eq.~(A21) in Ref.~\cite{Sobczyk:2021ejs}.

\end{widetext}

\end{document}